\documentclass[aps,pra,twocolumn,superscriptaddress,showpacs]{revtex4}
\usepackage{amssymb}
\usepackage{amsmath}
\usepackage{graphicx}
\usepackage{color}

\begin{document}

\title{Single- and double-slit collimating effects on fast-atom diffraction
spectra.}
\author{M.S. Gravielle\thanks{%
Author to whom correspondence should be addressed.\newline
Electronic address: msilvia@iafe.uba.ar}}
\affiliation{Instituto de Astronom\'{\i}a y F\'{\i}sica del Espacio (IAFE, CONICET-UBA),
casilla de correo 67, sucursal 28, C1428EGA, Buenos Aires, Argentina.}
\author{J.E. Miraglia}
\affiliation{Instituto de Astronom\'{\i}a y F\'{\i}sica del Espacio (IAFE, CONICET-UBA),
casilla de correo 67, sucursal 28, C1428EGA, Buenos Aires, Argentina.}
\date{\today }

\begin{abstract}
Diffraction patterns produced by fast He atoms grazingly impinging on a
LiF(001) surface are investigated focusing on the influence of the beam
collimation. Single- and double- slit collimating devices situated in front
of the beam source are considered. To describe the scattering process we use
the Surface Initial Value Representation (SIVR) approximation, which is a
semi-quantum approach that incorporates a realistic description of the
initial wave packet in terms of the collimating parameters. Our initial
wave-packet model is based on the Van Cittert-Zernike theorem. For a single-
slit collimation the width of the collimating aperture controls the shape of
the azimuthal angle distribution, making different interference mechanisms
visible, while the length of the slit affects the polar angle distribution.
Additionally, we found that by means of a double-slit collimation it might
be possible to obtain a wide polar angle distribution, which is associated
with a large spread of the initial momentum perpendicular to the surface,
derived from the uncertainty principle. It might be used as a simple way to
probe the surface potential for different normal distances.
\end{abstract}

\maketitle

\section{Introduction}

In the last time grazing-incidence fast atom diffraction (GIFAD or FAD) \cite%
{Schuller07,Rousseau07} has emerged as a powerful surface analysis technique
that allows one to inspect ordered surfaces, providing detailed information
on their morphological and electronic characteristics \cite%
{Winter11,Debiossac14,Zugarramurdi15}. The extreme sensitivity of FAD
patterns to the projectile-surface interaction relies on the preservation of
quantum coherence \cite{Aigner08,Lienemann11,Bundaleski11,Rubiano14}, and in
this aspect the collimating conditions of the incident beam play an
important role. In particular, recent experimental \cite{Winter15} and
theoretical \cite{Gravielle15} works have shown that FAD patterns are
strongly affected by the width of the collimating aperture, which determines
that two different mechanisms - Bragg diffraction or supernumerary rainbows
- can be alternatively observed.

In this article we theoretically investigate the influence of beam
collimation on FAD spectra by considering not only a single-slit but also a
double-slit \ collimating device. In the model, the transversal \textit{%
coherence} size of the initial wave packet associated with the incident
particle is determined from the collimating conditions by means of the Van
Cittert-Zernike theorem \cite{BornWolf}, while the elastic atom-surface
scattering is described within a recently developed semi-quantum approach,
named Surface-Initial Value Representation (SIVR) approximation \cite%
{Gravielle14}. \ The SIVR method offers a clear representation of the main
mechanisms of the process in terms of classical trajectories through the
Feynman path integral formulation of quantum mechanics \cite{Miller01}. It
includes an approximate description of classically forbidden transitions on
the dark side of the rainbow angle, giving a successful representation of
the experimental FAD patterns over the whole angular range \cite%
{Gravielle14,Gravielle15}, without requiring the use of convolutions to
smooth the theoretical curves \cite{Rubiano13}.

The SIVR approximation is here applied to evaluate FAD patterns for He atoms
grazingly impinging on a LiF(001) surface after going through a collimating
aperture formed by one or two parallel slits. The size and separation of
such slits, which limit the effective size of the extended incoherent beam
source, determine the aspect of FAD patterns, making different interference
mechanisms visible. The paper is organized as follows. The theoretical
formalism is summarized in Sec. II. Results for different sizes of the
collimating apertures are presented and discussed in Sec. III, while in Sec.
IV we outline our conclusions. Atomic units (a.u.) are used unless otherwise
stated.

\section{Theoretical model}

We consider an atomic projectile ($P$), with initial momentum $\vec{K}_{i}$,
which is elastically scattered from a crystal surface ($S$), ending in a
final state with momentum $\vec{K}_{f}$ \ and total energy $%
E=K_{f}^{2}/(2m_{P})=K_{i}^{2}/(2m_{P})$, $m_{P}$ being the projectile mass.
The frame of reference is located on the first atomic layer, with the
surface contained in the $x-y$ plane, the $\widehat{x}$ versor along the
incidence direction and the $\hat{z}$ versor oriented perpendicular to the
surface, aiming towards the vacuum region (see Fig. 1).

Within the SIVR approximation \cite{Gravielle14}, the amplitude per unit of
surface area $\mathcal{S}$ for the transition $\vec{K}_{i}\rightarrow \vec{K}%
_{f}$\ \ reads
\begin{eqnarray}
A_{if}^{{\small (SIVR)}} &=&\frac{1}{\mathcal{S}}\int\limits_{\mathcal{S}}d%
\overrightarrow{R}_{o}\ f_{i}(\overrightarrow{R}_{o})\int d\overrightarrow{K}%
_{o}\ g_{i}(\overrightarrow{K}_{o})  \notag \\
&&\times \ a_{if}^{{\small (SIVR)}}(\overrightarrow{R}_{o},\overrightarrow{K}%
_{o}),  \label{Aif-sivr}
\end{eqnarray}%
where $a_{if}^{{\small (SIVR)}}(\overrightarrow{R}_{o},\overrightarrow{K}%
_{o})$ is the partial transition amplitude associated with the classical
projectile path $\mathcal{\vec{R}}_{t}\equiv \mathcal{\vec{R}}_{t}(%
\overrightarrow{R}_{o},\overrightarrow{K}_{o})$, with $\overrightarrow{R}%
_{o} $ and $\overrightarrow{K}_{o}$ being the starting position and
momentum, respectively, at the time $t=0$. The functions $f_{i}(%
\overrightarrow{R}_{o}) $ and $g_{i}(\overrightarrow{K}_{o})$ describe the
profiles of the position and momentum distributions of the initial wave
packet, which depend on the beam collimation.

In Eq. (\ref{Aif-sivr}) the starting position can be expressed as $%
\overrightarrow{R}_{o}=\overrightarrow{R}_{os}+Z_{o}\widehat{z}$, where $%
\overrightarrow{R}_{os}=$ $X_{o}\widehat{x}+Y_{o}\widehat{y}$ is the
component parallel to the surface plane and $Z_{o}$ is a fixed distance,
here chosen as equal to the lattice constant, for which the projectile is
hardly affected by the surface interaction. The partial transition amplitude
$a_{if}^{{\small (SIVR)}}$ reads
\begin{eqnarray}
a_{if}^{{\small (SIVR)}}(\overrightarrow{R}_{o},\overrightarrow{K}_{o}) &=&\
-\int\limits_{0}^{+\infty }dt\ \ \frac{\left\vert J_{M}(t)\right\vert
^{1/2}e^{i\nu _{t}\pi /2}}{(2\pi i)^{9/2}}V_{SP}(\mathcal{\vec{R}}_{t})
\notag \\
&&\times \exp \left[ i\left( \varphi _{t}^{{\small (SIVR)}}-\overrightarrow{Q%
}\cdot \overrightarrow{R}_{o}\right) \right] ,  \label{aif}
\end{eqnarray}%
where $V_{SP}$ represents the surface-projectile interaction, $%
\overrightarrow{Q}=\vec{K}_{f}-\vec{K}_{i}$ is the projectile momentum
transfer, and
\begin{equation}
\varphi _{t}^{{\small (SIVR)}}=\int\limits_{0}^{t}dt^{\prime }\ \left[ \frac{%
1}{2m_{P}}\left( \vec{K}_{f}-\overrightarrow{\mathcal{P}}_{t^{\prime
}}\right) ^{2}-V_{SP}(\mathcal{\vec{R}}_{t^{\prime }})\right]  \label{fitot}
\end{equation}%
is the SIVR phase at the time $t$, with $\overrightarrow{\mathcal{P}}%
_{t}=m_{P}d\mathcal{\vec{R}}_{t}/dt\ $ the classical projectile momentum. \
In Eq. (\ref{aif}) the Maslov function \cite{Guantes04}
\begin{equation}
J_{{\small M}}(t)=\det \left[ \frac{\partial \mathcal{\vec{R}}_{t}(%
\overrightarrow{R}_{o},\overrightarrow{K}_{o})}{\partial \overrightarrow{K}%
_{o}}\right] =\left\vert J_{M}(t)\right\vert \exp (i\nu _{t}\pi )  \label{J}
\end{equation}%
is a Jacobian factor (a determinant) evaluated along the classical
trajectory $\mathcal{\vec{R}}_{t}$, with $\left\vert J_{{\small M}%
}(t)\right\vert $ the modulus of $J_{M}(t)$ and $\nu _{t}$ an integer number
that accounts for the sign of $J_{M}(t)$ at a given time $t$, satisfying
that every time that $J_{{\small M}}(t)$ changes its sign along the
trajectory, $\nu _{t}$ increases by $1$.

In this work we consider a collimating device formed by $n$ equivalent
rectangular apertures (with $n=1$ or $2$) placed in front of an extended
incoherent beam source. Each of the rectangular openings is oriented in such
a way that the corresponding transversal width, $d_{y}$, is parallel to the
surface (i.e. parallel to the $\widehat{y}$ versor), while the side of
length $d_{x}$ forms an angle $\theta _{x}=\pi /2-\theta _{i}$ with the
surface (i.e. with the $\widehat{x}$ versor), with $\theta _{i}$ being the
glancing incidence angle, as depicted in Fig. 1. We assume that the spatial
profile of the coherent initial wave packet at a distance $Z_{o}$ from the
surface is determined by the complex degree of coherence $\mu
_{n}(X_{o},Y_{o})$ \cite{BornWolf,Schaff14}. Extending the Van
Cittert-Zernike theorem \cite{BornWolf} to deal with an atomic beam passing
through the collimating opening, $\mu _{n}(X_{o},Y_{o})$ can be expressed as
\begin{eqnarray}
\left\vert \mu _{n}(X_{o},Y_{o})\right\vert ^{2} &=&j_{0}^{2}(\frac{\pi d_{x}%
}{\lambda _{\bot }L_{c}}X_{o})j_{0}^{2}(\frac{\pi d_{y}}{\lambda L_{c}}Y_{o})
\label{ucoh} \\
&&\times \cos ^{2}((n-1)\frac{\pi b}{\lambda _{\bot }L_{c}}X_{o}),\quad
\notag
\end{eqnarray}%
where $n=1,2$, represents the number of collimating slits, $L_{c}$ \ is the
collimator-surface distance, $b$ is the distance between the centers of the
slits, and $j_{0}(x)$ is the spherical Bessel function. The de Broglie
wavelengths $\lambda $ and $\lambda _{\bot }$ are defined as
\begin{equation}
\lambda =2\pi /K_{i}\text{,}\;\text{and }\;\lambda _{\bot }=\lambda /\sin
\theta _{i},  \label{lambda}
\end{equation}%
respectively, this last one being associated with the initial motion normal
to the surface plane. At this point it should be mentioned that Eq. (\ref%
{ucoh}) represents a limit case of a more rigorous expression, given by Eq. (%
\ref{nu-0}), whose calculation involves a numerical integration. Details of
its derivation are given in the Appendix.

For small $X_{o}$ and $Y_{o}$\ values, the spatial profile of the initial
wave packet $\left\vert f_{i}(\overrightarrow{R}_{os})\right\vert ^{2}\simeq
\left\vert \mu _{n}(X_{o},Y_{o})\right\vert ^{2}$ can be approximated as a
product of Gaussian functions, $G(\omega ,x)=[2/(\pi \omega ^{2})]^{1/4}\exp
(-x^{2}/\omega ^{2})$, as%
\begin{equation}
f_{i}(\overrightarrow{R}_{os})=G(\sigma _{x},X_{o})G(\sigma _{y},Y_{o}),
\label{ffin}
\end{equation}%
where the parameters $\sigma _{x}$ and \ $\sigma _{y}$ determine the \textit{%
transversal} \textit{coherence} \textit{size} of the initial wave packet
\cite{Tonomura86}. Such parameters were obtained from a numerical fitting,
reading
\begin{equation}
\sigma _{x}=\frac{\lambda _{\bot }}{\sqrt{2}}\frac{L_{c}}{D_{x}(n)}%
,\;\;\sigma _{y}=\frac{\lambda }{\sqrt{2}}\frac{L_{c}}{d_{y}},
\label{sigmax}
\end{equation}%
with $D_{x}(1)=d_{x}$ for a single-slit collimator and $D_{x}(2)=b$ for a
double-slit one.

Concerning the starting momentum $\overrightarrow{K}_{o}$, since we are
dealing with an incident beam with a well defined energy, i.e., $\Delta
E/E\ll 1$ \cite{Winter15}, it satisfies energy conservation, with $%
K_{0}=\left\vert \vec{K}_{0}\right\vert =\sqrt{2m_{P}E}$. Therefore, the
momentum profile of the initial wave packet, $g_{i}(\overrightarrow{K}_{o})$%
, can be replaced by the angular profile
\begin{equation}
g_{i}(\overrightarrow{K}_{o})\simeq g_{i}(\Omega _{o})=G(\sigma _{\theta
},\theta _{o}-\theta _{i})G(\sigma _{\varphi },\varphi _{o}),  \label{ggi}
\end{equation}%
where $\Omega _{o}\equiv (\theta _{o},\varphi _{o})$ is the solid angle
corresponding to the $\overrightarrow{K}_{o}$ direction. The angular widths
of the $\theta _{o}$- and $\varphi _{o}$- distributions are derived from Eq.
(\ref{ffin}) by applying the Heisenberg uncertainty relation \cite{Cohen},
reading
\begin{equation}
\sigma _{\theta }=\frac{\lambda _{\bot }}{2\sigma _{x}}\text{, and}\;\sigma
_{\varphi }=\frac{\lambda }{2\sigma _{y}},  \label{sigmatita}
\end{equation}%
respectively.

Finally, by replacing Eqs. (\ref{ffin}) and (\ref{ggi}) in Eq. (\ref%
{Aif-sivr}),\ the SIVR\ transition amplitude can be expressed as
\begin{eqnarray}
A_{if}^{{\small (SIVR)}} &=&\ \frac{m_{P}K_{i}}{\mathcal{S}}\int\limits_{%
\mathcal{S}}d\overrightarrow{R}_{os}\ f_{i}(\overrightarrow{R}_{os})\int
d\Omega _{o}\ g_{i}(\Omega _{o})  \notag \\
&&\times a_{if}^{{\small (SIVR)}}(\overrightarrow{R}_{o},\overrightarrow{K}%
_{o}),  \label{Aif-sivrn}
\end{eqnarray}%
where $a_{if}^{{\small (SIVR)}}$ is given by Eq. (\ref{aif}). \ Details of
the derivation of the SIVR method are given in Refs. \cite%
{Gravielle14,Gravielle15}.

\section{Results}

With the aim of studying the dependence of FAD patterns on the collimation
conditions we apply the SIVR method to 1 keV $^{4}$He atoms impinging on a
LiF(001) surface along the $\left\langle 110\right\rangle $ channel, with
the incidence angle $\theta _{i}=0.99$ deg. For this collision system,
experimental results obtained by using a single collimating aperture\ with
different widths were reported in Ref. \cite{Winter15}.

In this section, results for single-slit and double-slit collimating devices
situated in front of an extended incoherent beam source, at a distance $%
L_{c}=25$ cm from the surface \cite{Winter15}, will be separately
analyzed. The size of the beam source and its distance to the
collimator were chosen as $e\simeq 1$ cm and $L_{e}\simeq 100$ cm,
respectively, falling within the range where Eq. (\ref{ucoh}) is
valid, as given by Eq. (\ref{seda-lim}) \cite{nota}. We emphasize
that different collimation conditions can modify the coherence
lengths defined by Eqs. (\ref{sigmax}) and (\ref{sigmatita}),
affecting our results. In a more general case, the transversal
coherence size should be derived from the rigorous calculation of
Eq. (\ref{nu-0}).

The SIVR differential probability for elastic scattering with final momentum
$\vec{K}_{f}$ in the direction of the solid angle $\Omega _{f}\equiv (\theta
_{f},\varphi _{f})$ was derived from Eq. (\ref{Aif-sivrn}) as \cite%
{Gravielle14}
\begin{equation}
dP^{{\small (SIVR)}}/d\Omega _{f}=K_{f}^{2}\left\vert A_{if}^{{\small (SIVR)}%
}\right\vert ^{2},  \label{dPdangle}
\end{equation}%
with $\theta _{f}$ being the final polar angle, measured with respect to the
surface, and $\varphi _{f}$ being the azimuthal angle, measured with respect
to the $\widehat{x}$ axis (Fig. 1). The transition amplitude $A_{if}^{%
{\small (SIVR)}}$ was obtained from Eq. (\ref{Aif-sivrn}) \ by employing the
MonteCarlo technique with more than $4\times 10^{5}$ points in the $%
\overrightarrow{R}_{os}-$ and $\Omega _{o}-$ integrations. Like in Refs.
\cite{Gravielle14,Gravielle15}, the projectile-surface interaction was
evaluated with a pairwise additive potential that includes no local terms of
the electronic density in the kinetic and exchange contributions, as well as
projectile polarization and rumpling effects.

\subsection{Single collimating slit}

We start analyzing the influence of the width of a single collimating
aperture, $d_{y}$, which determines the effective length of the extended
beam source in the direction transversal to the incidence channel. In Fig. 2
we show two-dimensional projectile distributions, as a function of $\theta
_{f}$ and $\varphi _{f}$, derived within the SIVR approximation by
considering single collimating slits with the same length - $d_{x}=1.5$ mm -
but different $d_{y}$. \ For the narrowest aperture - $d_{y}=0.1$ mm - the
final angular distribution presents thin peaks associated with Bragg
diffraction, which are situated at azimuthal angles that verify \cite%
{Schuller09}
\begin{equation}
\sin \varphi _{f}=m\lambda /a_{y},  \label{Bragg}
\end{equation}%
where $m$ is an integer\ number that determines the Bragg order and $a_{y}$\
is the length of the reduced unit cell along the $\widehat{y}$ direction.
From Fig. 2 we observe that the width of these Bragg peaks notoriously
increases as $d_{y}$ augments. In particular, for $d_{y}=0.6$ mm, Bragg
interference is almost completely blurred out and intense rainbow maxima at
the outermost azimuthal angles of the angular spectrum arise \cite%
{Gravielle08}. For wider collimating apertures, i.e., $d_{y}\gtrapprox 0.8$
mm, Bragg peaks fade out, giving way to supernumerary rainbow structures in
the final projectile distribution. As discussed in Refs. \cite%
{Winter15,Gravielle15}, this behavior is associated with the transversal
length $\mathcal{D}_{y}$ of the area $\mathcal{S}$ of the surface plane that
is coherently illuminated by the incident beam. The transversal length of
the coherently illuminated region can be estimated from Eq. (\ref{ucoh}) as $%
\mathcal{D}_{y}=2\sqrt{2}\sigma _{y}$ \cite{Gravielle15}, being inversely
proportional to $d_{y}$, as given by Eq. (\ref{sigmax}). Hence, for the
narrower collimating apertures of Fig. 2 several reduced unit cells in the
direction transversal to the incidence channel become coherently illuminated
by the initial wave packet, giving rise to sharp Bragg maxima. But when the
width of the collimating opening augments, the number of reduced unit cells
that are coherently lighted decreases, and for $d_{y}\gtrapprox 0.8$ mm only
one reduced unit cell is coherently illuminated along the $\widehat{y}$
direction. Consequently, interferences coming from different parallel
channels, which are associated with the Bragg mechanism \cite%
{Schuller09,SchullerGrav09}, disappear and the unit-cell form factor
corresponding to the interference inside one channel governs the projectile
distribution \cite{SchullerGrav09}, causing supernumerary rainbow structures
to be visible instead \cite{Schuller08}.

Such an effect of the width of the collimating aperture on FAD\ patterns was
also experimentally observed \cite{Winter15}. As shown in Figs. 2 and 3 of
Ref. \cite{Gravielle15}, SIVR diffraction spectra for two different
collimating widths - $d_{y}=0.2$ mm and $d_{y}=1.0$ mm \ - compare very well
with the corresponding experimental data extracted from Ref. \cite{Winter15}%
. For the narrowest aperture (Fig. 2 of Ref. \cite{Gravielle15}), the
experimental and theoretical distributions present well-defined Bragg peaks
laying on a thick annulus, whose mean radius is approximately equal to $%
\theta _{i}$. However, these interference structures completely vanish when
the width of the opening increases, as it happens in Fig. 3 of Ref. \cite%
{Gravielle15}, where only maxima at the rainbow deflection angles are
clearly observed.

In the case of a single collimating slit, the thickness of the annulus
corresponding to the $(\theta _{f},\varphi _{f})$ distribution of scattered
projectiles is controlled by the length of the collimating aperture,
increasing as $D_{x}(1)=d_{x}$ augments. In Fig. 3 we show angular
projectile distributions derived from the SIVR approach by considering \
single collimating slits with the same width - $d_{y}=0.2$ mm - and
different lengths. For the small square aperture of Fig. 3 (a), with $%
d_{x}=d_{y}=0.2$ mm, the Bragg peaks look like circular spots lying on a
thin ring whose radius is equal to $\theta _{i}$. But the thickness of this
ring augments as $d_{x}$ increases, and for $d_{x}=0.5$ mm (Fig. 3 (b)) the
spots become slightly elongated strips. For longer collimating apertures,
with $d_{x}\gtrapprox 1.5$ mm, not only the length of the interference
fringes augments but also the different Bragg orders are situated at
different radius, sampling additional interference structures along the $%
\theta _{f}$- axis. This effect is related to the $\theta _{o}$- spread of
the initial wave packet, which is determined by $\sigma _{\theta }$, being
proportional to $d_{x}$ as given by Eqs. (\ref{sigmatita}) and (\ref{sigmax}%
). Large $d_{x}$ values produce a wide spread of the impact momentum normal
to the surface plane, $\left\vert K_{oz}\right\vert =K_{o}\sin \theta _{o}$,
revealing interference structures similar to those observed in the
diffraction charts for different normal energies $E_{\bot }=E\sin ^{2}\theta
_{i}$. Then, the intensity oscillations along the $\theta _{f}$- axis
observed for long collimating openings are exploring the surface potential
for different distances to the topmost atomic plane.

\subsection{Double collimating slit}

Within the limits where Eq. (\ref{ucoh}) holds, for double collimating slits
the length of the coherently illuminated region along the incidence channel
is governed by $D_{x}(2)=b$, instead of $d_{x}$. Therefore, under similar
collimating conditions a double-slit collimating device produces a larger $%
\theta _{o}$- dispersion than the one corresponding to a single slit. Since
a wide $\theta _{o}$- spread leads to a large dispersion of the
perpendicular momentum $K_{oz}$, the angular projectile distributions
derived by using a double aperture with a separation $b$ of a few
millimeters display interference structures in the final polar angle axis,
as observed in Fig. 4. At the point that $b=$\ $5.0$ mm, the corresponding
angular distribution resembles the usual diffraction chart, which is
obtained with a single-slit collimator by varying the incidence angle $%
\theta _{i}$, or what is the same, the normal energy $E_{\bot }$, to cover
the whole perpendicular energy spectrum. In Fig. 5 we compare the SIVR
angular spectrum derived by using a double-slit collimating device with $b=$%
\ $5.0$ mm with the diffraction chart obtained from the SIVR approach with a
single slit, both collimating apertures with the same size: $d_{x}=d_{y}=0.2$
mm. The similarities between the projectile distributions of Figs. 5 (a) and
5 (b) are evident, reinforcing the idea that double collimating slits might
be employed to probe potential energy surfaces corresponding to different
normal energies, instead of the commonly used diffraction charts. Similar
angular distributions can be also obtained by employing a single slit with a
sufficiently large $d_{x}$ height.

\smallskip

Lastly, notice that any change in the present collimation setups might
affect our results and it should be specifically investigated by using a
generalization of Eq. (\ref{nu-0}). Moreover, \ we have considered fixed
incidence conditions, i.e., constant de Broglie wavelengths. However, the
spatial spread of the initial wave packet depends on $\lambda $ and $\lambda
_{\bot }$, as given by Eq. (\ref{sigmax}). Then, a variation of the
incidence energy or angle would modify such a spread, affecting the shape of
the diffraction patterns. Also, the presence of crystal defects might play a
role contributing to additional angular dispersions.

\section{Conclusions}

We have investigated the effect of the collimation of the incident beam on
FAD patterns by considering a single- and \ a double- slit collimator that
limits the effective size of an extended incoherent atom source. Projectile
distributions originated by elastic scattering were derived from the SIVR\
approximation \cite{Gravielle14} by incorporating a realistic description of
the coherent initial wave packet in terms of the collimating parameters. The
profile of the initial wave packet was derived from a model based on the Van
Cittert-Zernike theorem, which relates the effective size of the beam source
with the complex degree of coherence. The theory was applied to helium atoms
grazingly impinging on a LiF(001) surface, considering a\ fixed incidence
condition and different sizes of the rectangular collimating apertures.

In this work we found that the collimating effects on the final polar and
azimuthal distributions are decoupled. While the $\varphi _{f}$%
--distribution is affected by the transversal width of the collimating slit,
the $\theta _{f}$-- distribution is governed by the geometrical
characteristics of the collimating device along $x$. For a single-slit
collimator, the SIVR interference patterns are strongly affected by the
width of the collimating aperture, which makes Bragg peaks (supernumerary
rainbows) visible for narrow \ (wide) apertures. In turn, the length of the
collimating opening affects the polar angle distribution of scattered
projectiles, making visible additional intensity oscillations in the $\theta
_{f}$- axis for long collimating apertures. This effect is related to the
dispersion of the normal momentum $K_{oz}$ and it is more evident for a
double-slit collimator.

In the case of a double-slit collimating device, the interference patterns
along the final polar angle axis are governed by the separation between
slits. By using separation distances between slits of \ several millimeters
it would be possible to obtain projectile distributions covering a wide
range of final polar angles. Such distributions might be used to probe the
projectile-surface interaction for different normal distances. Finally, it
should be stressed that actual experimental setups can involve different
collimating stages, whose full description is necessary in order to obtain a
proper representation of the collimating effects.

\begin{acknowledgments}
The authors acknowledge financial support from CONICET, UBA, and ANPCyT of
Argentina.
\end{acknowledgments}

\appendix*

\section{Complex degree of coherence for an atomic beam passing through a
collimating aperture}

Here we extend the Van Cittert-Zernike theorem \cite{BornWolf} to evaluate
the complex degree of coherence for two points - $Y_{1}$ and $Y_{2}$ -
placed on a plane parallel to the crystal surface at a distance $Z_{o}$,
which is illuminated by an extended incoherent quasi-monochromatic source
after passing through a collimating aperture. For simplicity we study the
simplest two-dimensional case depicted in Fig. 6, where one-dimensional
emitter source and slit, with lengths $e$ \ and $d$ respectively, are
considered. The generalization for the three-dimensional case or for
two-slit collimating devices is straightforward.

As the source is composed by incoherent emitters we resort to the Van
Cittert-Zernike theorem \cite{BornWolf} to calculate the mutual intensity
function $J(Y_{1},Y_{2})$ as
\begin{equation}
J(Y_{1},Y_{2})=\int\limits_{-e/2}^{e/2}dy_{e}\
I_{o}\int\limits_{-d/2}^{d/2}dy_{1}\int\limits_{-d/2}^{d/2}dy_{2}\frac{\exp %
\left[ ik(R_{1}-R_{2})\right] }{r_{1}s_{1}r_{2}s_{2}},  \label{J12}
\end{equation}%
where $I_{o}$ is the intensity of the extended source, assumed as uniform, $%
k=2\pi /\lambda $ is the wave number of the atomic beam, and $%
R_{j}=s_{j}+r_{j}$, with the distances $s_{j}$ and $r_{j}$ being indicated
in Fig. 6 for $j=1,2$. By considering, as usually, that the distances $L_{e}$
and $L_{c}$ between the source and the collimator and between the
collimating slit and the upon surface plane, respectively, are larger than $e
$, $d$, and $\Delta =Y_{1}-Y_{2}$, the mutual intensity function can be
reduced, except a constant factor, to :
\begin{equation}
J(Y_{1},Y_{2})=\exp \left[ i\alpha _{o}(Y_{1},Y_{2})\right] \ g\left(
Y_{1},Y_{2}\right) ,  \label{J12red}
\end{equation}%
where
\begin{eqnarray}
g\left( Y_{1},Y_{2}\right)  &=&\int\limits_{-e/2}^{e/2}dy_{e}\ \exp \left[
-iy_{e}\beta _{o}(Y_{1},Y_{2})\right] \times   \notag \\
&&E(Y_{1},y_{e})E^{\ast }(Y_{2},y_{e})  \label{g12}
\end{eqnarray}%
involves a one-dimensional integral and the asterisk indicates the complex
conjugate. In Eq. (\ref{g12}), the function $E(Y_{j},y_{e})$ is defined as%
\begin{equation}
E(Y_{j},y_{e})=\left[ \mathcal{C}(u_{j}^{(+)})+i\mathcal{S}(u_{j}^{(+)})%
\right] -\left[ \mathcal{C}(u_{j}^{(-)})+i\mathcal{S}(u_{j}^{(-)})\right] ,
\label{Er}
\end{equation}%
for $j=1,2$, with $\mathcal{C}(u)$ and $\mathcal{S}(u)$ the cosine and sine
Fresnel integrals \cite{Abramowitz},
\begin{equation}
u_{j}^{(\pm )}=\sqrt{\frac{k}{\pi L}}\left( \pm \frac{d}{2}-\frac{L_{e}}{%
L_{tot}}Y_{j}-\frac{L_{c}}{L_{tot}}y_{e}\right) ,\text{ for }j=1,2\text{,}
\label{uj}
\end{equation}%
and $L_{tot}=L_{e}+L_{c}$. The parameters $\alpha _{o}$ and $\beta _{o}$ read%
\begin{eqnarray}
\alpha _{o}(Y_{1},Y_{2}) &=&k(Y_{1}^{2}-Y_{2}^{2})/\left( 2L_{tot}\right) ,
\notag \\
\beta _{o}(Y_{1},Y_{2}) &=&k(Y_{1}-Y_{2})/L_{tot}.  \label{alfbet}
\end{eqnarray}%
Finally, the complex degree of coherence between $Y_{1}$ and $Y_{2}$ is
obtained from Eq. (\ref{J12red}) as%
\begin{eqnarray}
\mu (Y_{1},Y_{2}) &=&\frac{J(Y_{1},Y_{2})}{\sqrt{J(Y_{1},Y_{1})J(Y_{2},Y_{2})%
}}=  \label{nu-p12n} \\
&=&\exp \left[ i\alpha _{o}(Y_{1},Y_{2})\right] \frac{g\left(
Y_{1},Y_{2}\right) }{\sqrt{g\left( Y_{1},Y_{1}\right) g\left(
Y_{2},Y_{2}\right) }},  \notag
\end{eqnarray}%
where the function $g\left( Y_{1},Y_{2}\right) $ is defined by Eq. (\ref{g12}%
).

\smallskip

In this article, in order to derive the profile of the incident wave packet
it is convenient to choose $Y_{1}=0$ as the center of the wave packet, while
$Y_{2}=Y_{o}$ represents the coherence distance. Hence, the square modulus
of the corresponding complex degree of coherence can be obtained from Eq. (%
\ref{nu-p12n}) as
\begin{equation}
\left\vert \mu (Y_{o})\right\vert ^{2}=\frac{\left\vert g\left(
0,Y_{o}\right) \right\vert ^{2}}{g\left( 0,0\right) g\left(
Y_{o},Y_{o}\right) }.  \label{nu-0}
\end{equation}%
The calculation of the complex degree of coherence from Eq. (\ref{nu-0})
requires the numerical evaluation of the integral given by Eq. (\ref{g12}).
However, analytical expressions can be obtained by considering two different
mathematical limits; that is%
\begin{equation}
\left\vert \mu (Y_{o})\right\vert ^{2}\approx j_{0}^{2}(\xi \frac{k}{2}%
Y_{o}),  \label{nu-1a}
\end{equation}%
where%
\begin{equation}
\xi =\left\{
\begin{array}{cc}
e/L_{tot},\text{\ \ } & \text{if \ }e\ll L_{tot}\ d/L_{c} \\
d/L_{c}, & \text{if \ }e\gg L_{e}/(kd)%
\end{array}%
\right. .  \label{seda-lim}
\end{equation}%
Then, if the coherence size of the initial wave packet depends on $d$, as
found in Ref. \cite{Winter15}, \ from Eq. (\ref{seda-lim}) we expect that
the \textit{effective} size of the source was comparatively large (in our
case, $e\simeq $ 1 cm) and consequently, the distance $L_{c}$ had to be
considered in the $\xi $ definition. \ For other experimental setups, the
numerical calculation of Eq. (\ref{nu-0}) should be carried out in order to
estimate the value of the parameter $\xi $.

\begin{figure}[tbp]
\includegraphics[width=0.4\textwidth]{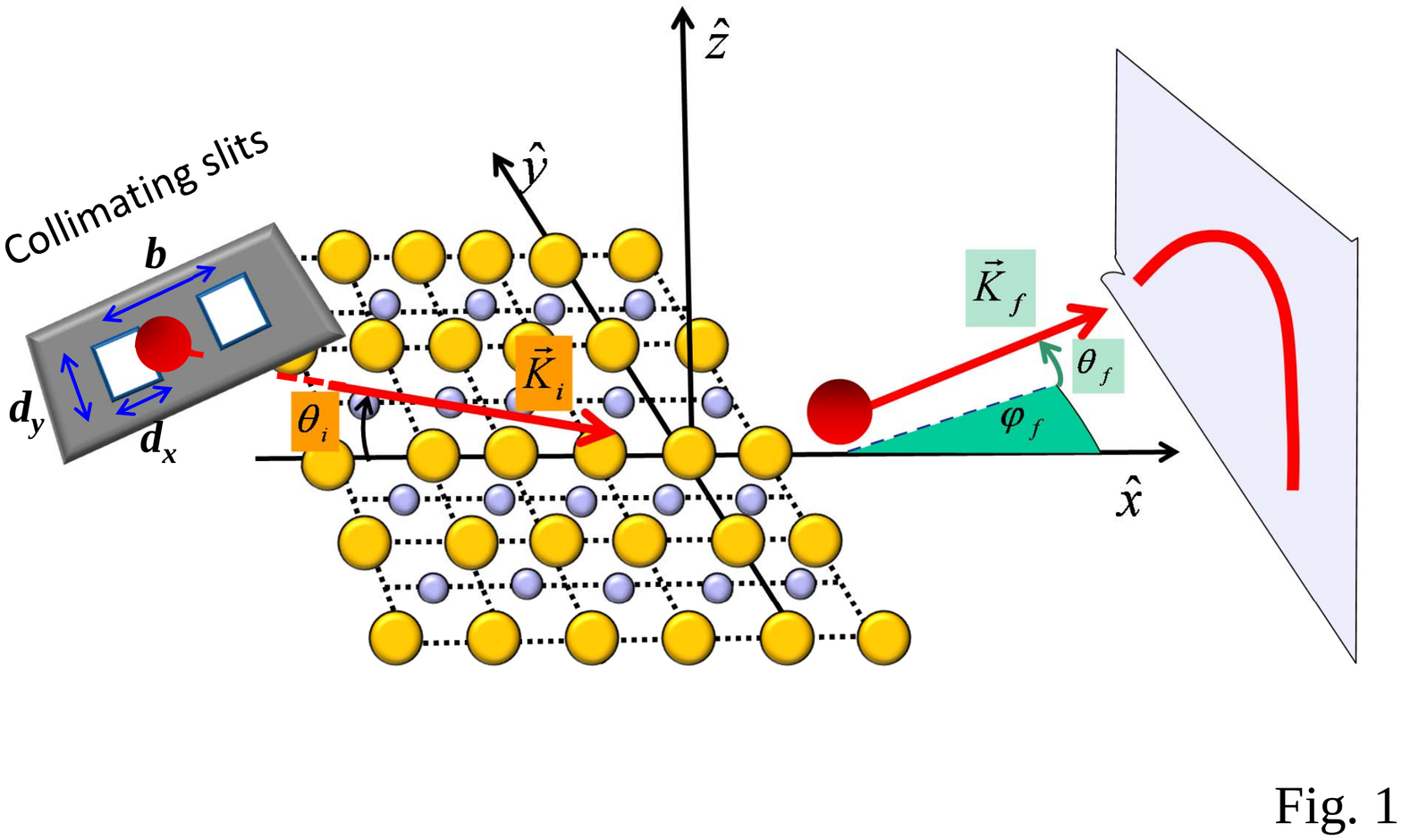}
\caption{(Color online) Depiction of the FAD process, including the
collimating device and the frame of reference. }
\label{Fig1}
\end{figure}

\begin{figure}[tbp]
\includegraphics[width=0.4\textwidth]{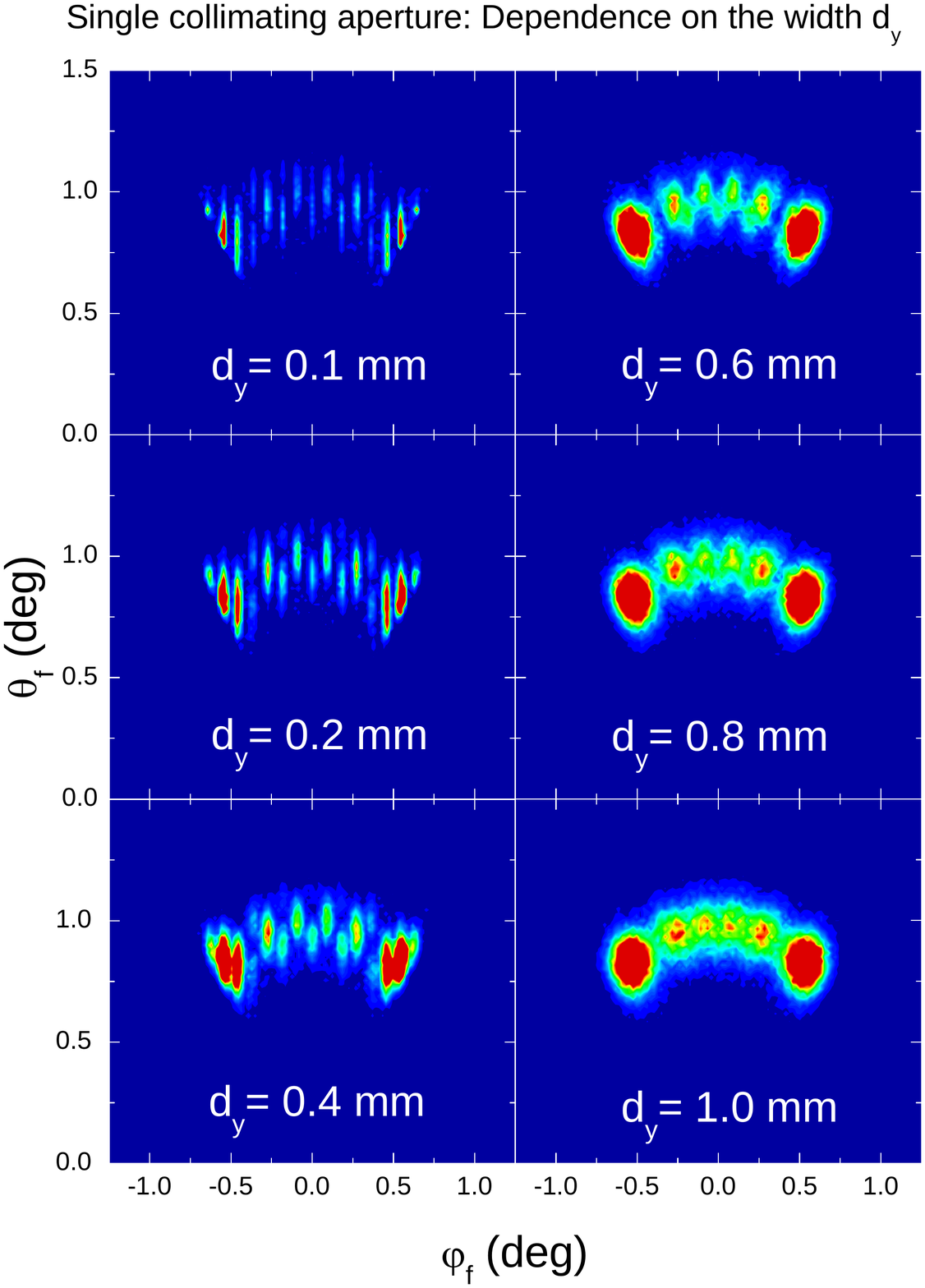}
\caption{(Color online) Two-dimensional projectile distribution, as a
function of the final dispersion angles $\protect\theta _{f}$ and $\protect%
\varphi _{f}$ , for $1$ keV $^{4}$He atoms impinging on LiF(001) along the $%
\left\langle 110\right\rangle $ direction with the incidence angle $\protect%
\theta _{i}=0.99$ deg. The helium beam is collimated with a single
rectangular aperture of length $d_{x}=$ \ $1.5$ mm and different widths: $%
d_{y}=$ \ $0.1$, $0.2$, $0.4$, $0.6$, $0.8$, and $1.0$ mm. }
\label{Fig2}
\end{figure}

\begin{figure}[tbp]
\includegraphics[width=0.4\textwidth]{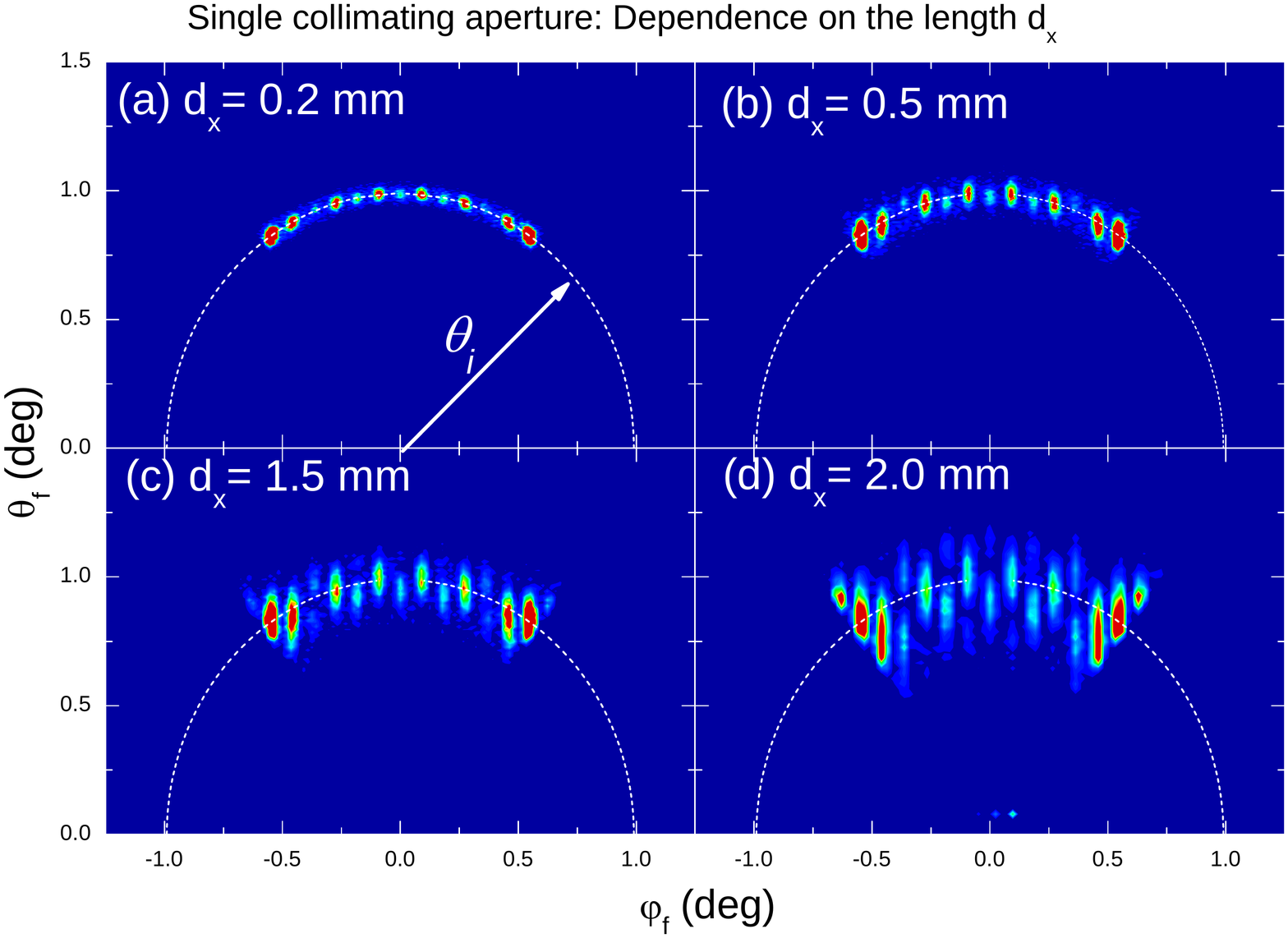}
\caption{(Color online) Similar to Fig. 2 for a single-slit collimating
device of width $d_{y}=$ \ $0.2$ mm and different lengths: (a) $d_{x}=$ \ $%
0.2$ mm, (b) $d_{x}=$ \ $0.5$ mm, (c) $d_{x}=$ \ $1.5$ mm, and (d)
$d_{x}=$ \ $2.0$ mm. In all the panels: dashed line, circle of
radius equal to the incidence angle $\protect\theta _{i}$.}
\label{Fig3n}
\end{figure}

\begin{figure}[tbp]
\includegraphics[width=0.4\textwidth]{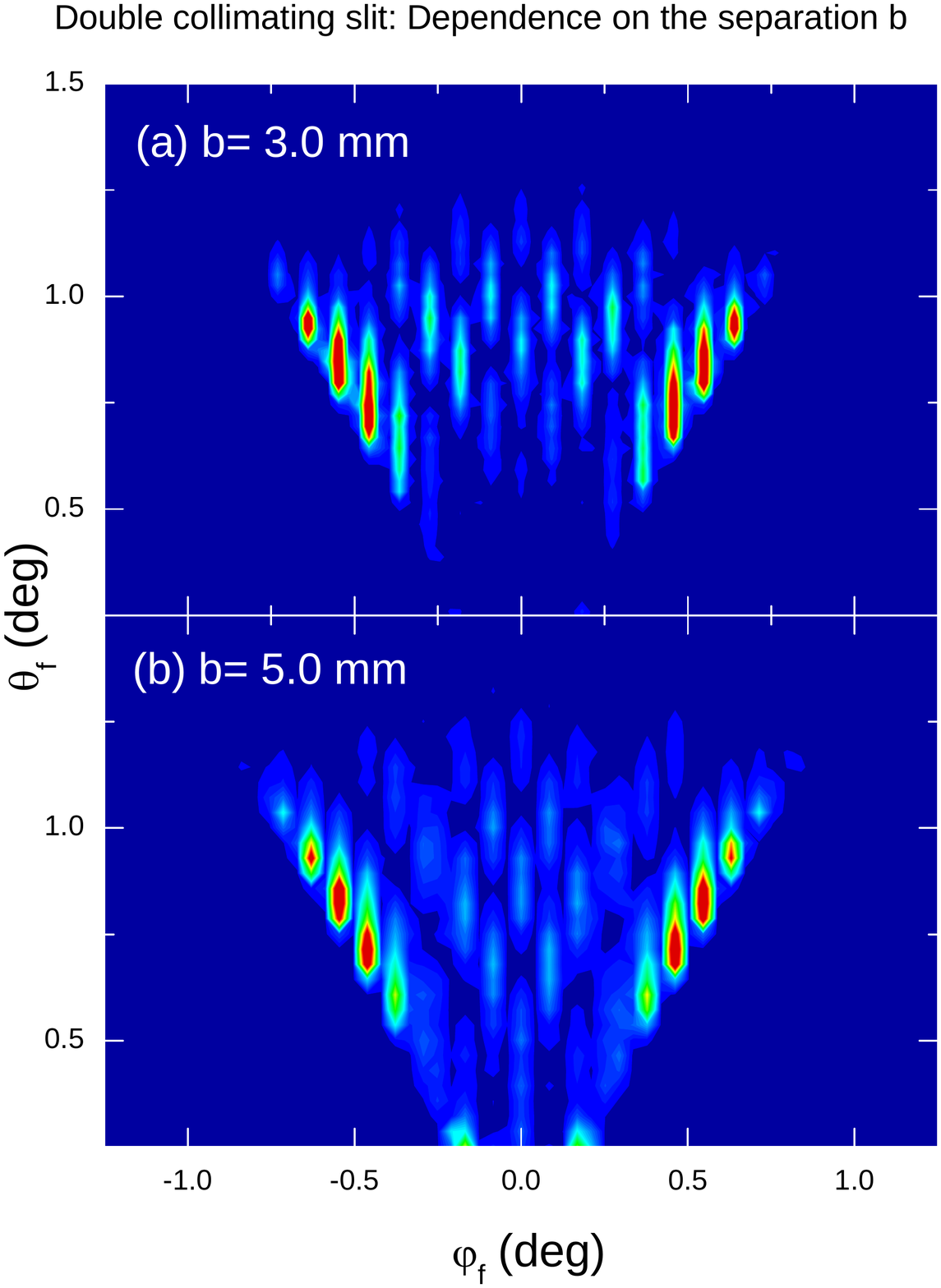}
\caption{(Color online) Similar to Fig. 2 for a double-slit
collimating device, each of the apertures of size $d_{x}=d_{y}=$ \
$0.2$ mm. The apertures are separated by a distance $b$, with: (a)
$b=$ \ $3.0$ mm , and (b) $b=$ \ $5.0$ mm.} \label{Fig4n}
\end{figure}

\begin{figure}[tbp]
\includegraphics[width=0.4\textwidth]{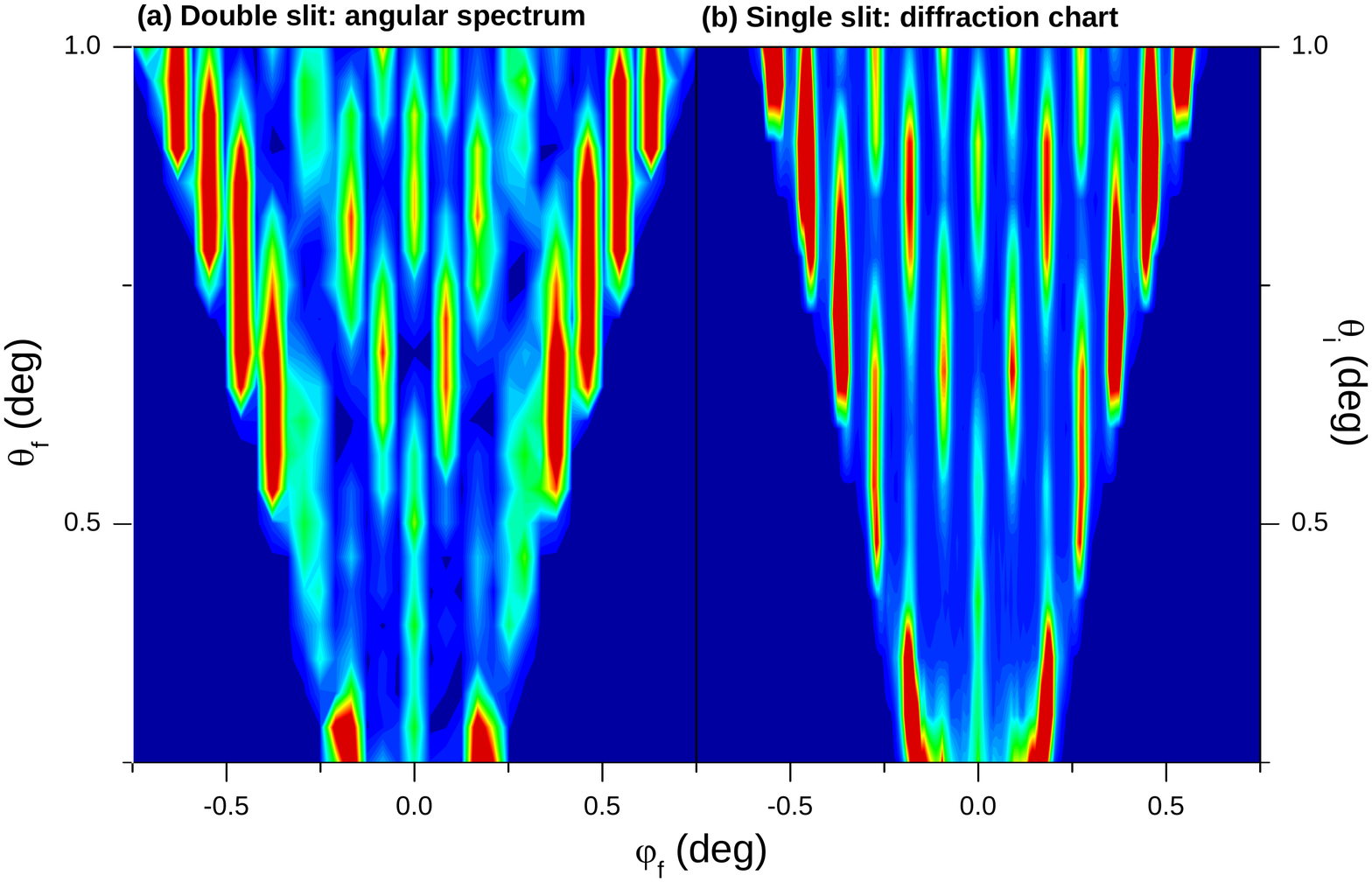}
\caption{(Color online) SIVR projectile distributions for $1$ keV
$^{4}$He atoms impinging on LiF(001) along the $\left\langle
110\right\rangle $ channel. (a) Angular distribution derived by
considering the incidence angle $\protect\theta _{i}=0.99$ deg and a
double-slit collimating device with a separation $b=$ \ $5.0$ mm.
(b) Diffraction chart derived by considering different incidence
angles $\protect\theta _{i}$ and a single-slit collimating device.
Both collimating apertures of size $d_{x}=d_{y}=$ \ $0.2$ mm.}
\label{Fig5n}
\end{figure}

\begin{figure}[tbp]
\includegraphics[width=0.4\textwidth]{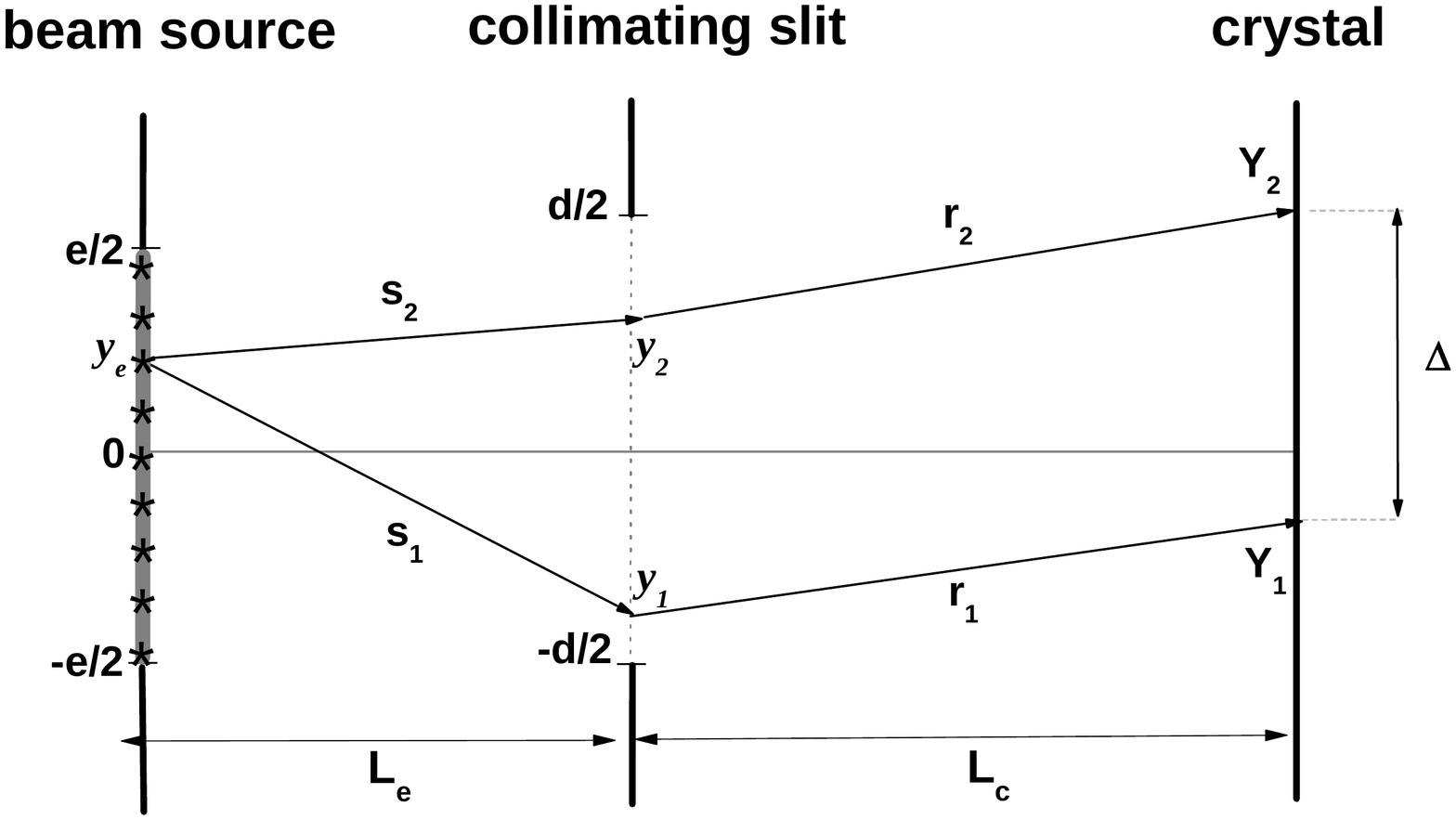}
\caption{Depiction of the two-dimensional system considered in the
Appendix, together with the involved coordinates.} \label{Fig6n}
\end{figure}


\begin{thebibliography}{99}
\bibitem{nota} Although such a large value of $e$ may be considered not in
accordance with realistic experimental setups, it might be also interpreted
as an effective source length due to the different collimation stages that
affect actual atomic beams.
\end{thebibliography}
\end{document}